\documentclass{kapedbk} 

\usepackage[dvips]{graphicx}
\usepackage{psfig}

 \chaptersection 

\upperandlowercase

\kluwerbib 

\newcommand{\gtwid}
           {\mathrel{\raise.25ex\hbox{$>$\kern-.80em\lower1ex\hbox{$\sim$}}}}
\newcommand{\Ha}{H$\alpha$}

\newcommand{\HII}{\sc H~ii\rm}

\newcommand{\CaII}{Ca~{\sc ii}}
\newcommand{\MgIb}{Mg~{\sc i}$b$} 
\newcommand{\NaID}{Na~{\sc i}~D}
\newcommand{\farcs}{\hbox{$.\!\!^{''}$}}

\begin{document}
\articletitle{Distant X-ray Galaxies: Insights\\
                    from the Local Population}

\rhead{Distant X-ray Galaxies}

\author{Edward C.\ Moran}

\affil{Astronomy Department \\
Wesleyan University, Middletown, CT 06459 USA}

\begin{abstract}
A full understanding of the origin of the hard X-ray background requires
a complete and accurate census of the distant galaxies that produce it.
Unfortunately, distant X-ray galaxies tend to be very faint at all
wavelengths, which hinders efforts to perform this census.  This chapter
discusses the insights that can be obtained through comparison of the
distant population to local X-ray galaxies, whose properties are well
characterized.  Such comparisons will ultimately aid investigations into
the cosmic evolution of supermassive black holes and their environments.  
\end{abstract}


\section{Introduction}

The sensitivities of the {\it Chandra\/} and {\it XMM-Newton X-ray 
Observatories\/} have made it possible, for the first time, to detect 
and study {\it typical\/} X-ray galaxies at cosmological distances.  
As a result, we are now able to conduct direct investigations of the 
sources that produce the hard ($>$~2~keV) X-ray background (XRB) 
radiation, whose origin remains one of the fundamental issues in 
X-ray astronomy.  Research over the past decade has indicated that
the majority of the hard XRB must arise from distant black hole 
powered Active Galactic Nuclei (AGN). Thus, we expect that the 
XRB---once it is fully resolved and the sources contributing to it 
are identified---will provide insight into the accretion histories 
of supermassive black holes and a record of their growth over cosmic 
time (Barger et al.\ 2001a).  Since the masses of supermassive black 
holes are linked to the properties of the galactic bulges within 
which they reside (see Ferrarese, this volume), this record could 
hold important clues about galaxy evolution in general.  One of our 
key near-term goals, therefore, is to obtain an accurate census of 
the distant X-ray galaxy population.  Unfortunately, the general 
faintness of this population presents an obstacle.  This chapter 
explores what can be learned about distant X-ray galaxies through 
comparison to the well-characterized local population.

\section{Seyfert~2 Galaxies and the X-ray Background}

It now appears that different types of sources produce the majority 
of the XRB in different X-ray energy bands.  In the soft ($0.5-2$~keV) 
band, AGN with broad optical emission lines (i.e., Seyfert~1
galaxies and quasars) are the dominant contributors 
(Schmidt et al.\ 1998).  These sources, however, have steep X-ray
spectra and cannot account for the flat slope of the XRB spectrum at 
higher energies (e.g., Mushotzky et al.\ 1980; Nandra \& Pounds 1994).  
Narrow line Seyfert~2 galaxies have thus emerged as the most 
promising candidates for the origin of the hard ($2-10$~keV) XRB.  
According to the unified AGN picture (Antonucci 1993), the soft X-ray 
fluxes of these sources are heavily absorbed by dense circumnuclear 
gas, presumably the same material that obscures their
broad emission-line regions.  As a result, their observed X-ray 
spectra are substantially flatter than those of unobscured AGN 
(Awaki et al.\ 1991). XRB models based on AGN unification have 
demonstrated that the combined emission of absorbed and unabsorbed 
sources, integrated over all redshifts, can account for the intensity 
and spectrum of the hard XRB (Setti \& Wolter 1989; Madau,
Ghisellini, \& Fabian 1993, 1994; Comastri et al.\ 1995).  According 
to these models, the sources responsible for most of the flux in 
the hard X-ray band are expected to have absorption column densities 
in the range $\log N_H\approx 23-24$, which are typical for Seyfert~2
galaxies (e.g., Risaliti, Maiolino, \& Salvati 1999).

The AGN based XRB models depend on a number of assumptions and free 
parameters that, in principle, can be verified or constrained 
observationally: (1) the universality of the unified AGN model 
(i.e., {\it all\/} Seyfert~2 galaxies are obscured Seyfert~1 galaxies), 
(2) the ratio of obscured to unobscured sources,
(3) the distribution and luminosity dependence of the absorption column
densities in AGN, and (4) the luminosity extent and redshift evolution of
the X-ray luminosity function for obscured AGN.  

Although there seems to
have been little debate in the X-ray community about the universality of
the unified AGN model, optical spectropolarimetry surveys, which can 
reveal broad emission lines in the polarized flux spectra of Seyfert~2
galaxies, have continued to the present, in part to test the unified model.
These surveys have demonstrated that at least $\sim$~50\% of such sources 
are indeed hidden Seyfert~1 galaxies (Moran et al.\ 2000, 2001; 
Lumsden et al.\ 2001; Tran 2001). 
Given the generally low polarizations of Seyfert~2 galaxies and 
other factors that make the spectropolarimetry technique difficult 
(e.g., Kay 1994), it is probably safe to assume that most Seyfert~2 
galaxies (at least those that produce the majority of the luminosity 
associated with this population) are obscured Seyfert~1 galaxies.
Improved constraints on some of the input model parameters, e.g., 
the column density distribution (Risaliti et al.\ 1999) and the 
evolution of the AGN X-ray luminosity function 
(Miyaji, Hasinger, \& Schmidt 2000), have permitted refinements
of the AGN model for the XRB (Risaliti et al.\ 1999;
Pompilio, La~Franca, \& Matt 2000; Gilli, Salvati, \& Hasinger 2001).

\begin{figure}[hbt]
\centerline{\psfig{figure=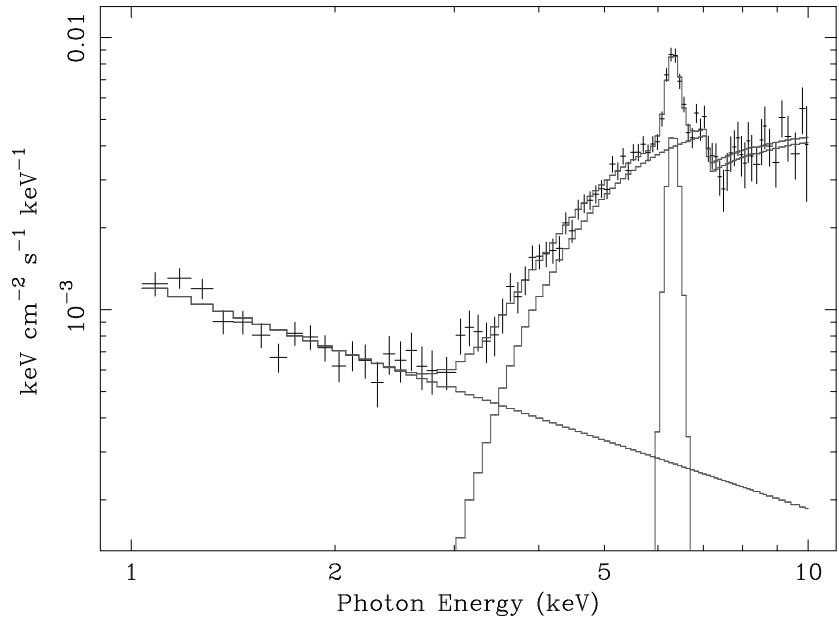,width=11truecm,angle=270}}
\caption{Composite Seyfert~2 X-ray spectrum derived from the
UW89 sample. The spectrum is fitted with a double power-law model
(see Moran et al.\ 2001). The composite spectrum confirms that
Seyfert~2 galaxies, as a class, have the spectral properties necessary
to explain the shape of the hard XRB spectrum.
}
\label{moranfig1}
\end{figure}

Concerned about the possible effects of selection biases on the 
observational inputs to the XRB models, my colleagues and I have 
adopted a different approach to investigating the contribution of 
Seyfert~2 galaxies to the hard XRB (Moran et al.\ 2001).  
Our study is based on the AGN sample of Ulvestad \& Wilson (1989,
hereafter UW89), which at the time of its definition included all known
Seyfert galaxies with recessional velocities $cz < 4600$ km~s$^{-1}$ and
declinations $\delta > -45^{\circ}$.  Because of its distance-limited nature,
this sample is free of serious selection biases.  Nearly all of the 31 
Seyfert~2 galaxies in the sample have been observed in the $1-10$~keV 
band with the {\em ASCA\/} satellite, which provides us with an 
opportunity to obtain a clear picture of the hard X-ray properties 
of these sources.  However, while all of the galaxies were detected 
with {\em ASCA\/}, many have proved to be weak X-ray sources.  
The data cannot be used, therefore, to measure the average
spectral parameters or column density distribution for the sample.  
Instead, we have combined the X-ray data for the sources in a 
luminosity-weighted fashion to obtain a composite Seyfert~2 spectrum.  
The result is displayed in Figure~\ref{moranfig1}.

%
%

\begin{figure}[tbh]
\centerline{\psfig{figure=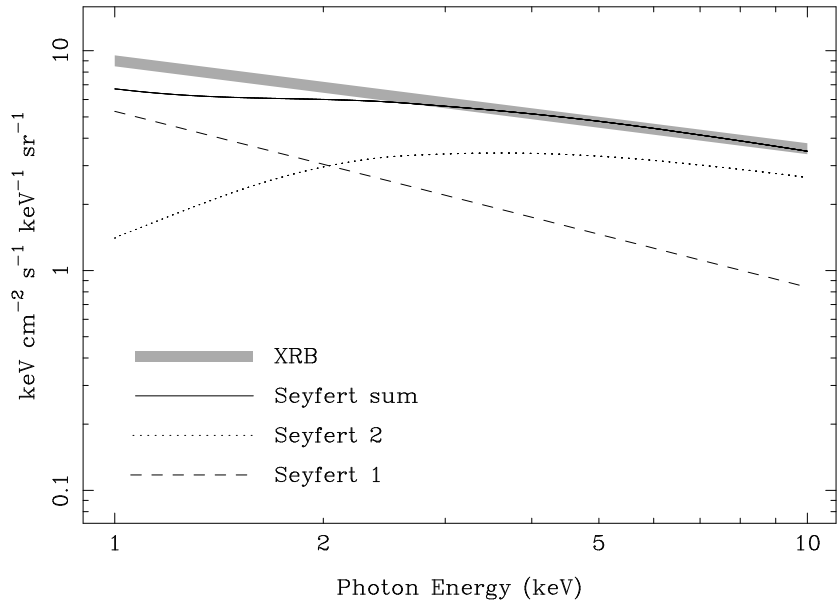,width=11truecm,angle=270}}
\caption{X-ray sky brightness due to Seyfert~2 galaxies,
obtained via integration of the composite spectrum in
Fig.~\ref{moranfig1} over redshift. The combined emission from
Seyfert~1 and Seyfert~2 galaxies compares well with the spectrum
of the XRB measured by Gendreau et al.\ (1995), especially at $E>3$~keV.
The discrepancy between the XRB spectrum and the Seyfert sum below
2~keV could represent the contributions of groups and clusters of galaxies
and star-forming galaxies, which are not included in this simple model.
}
\label{moranfig2}
\end{figure}

Because the sample from which it is derived is volume-limited, the composite
spectrum is effectively a measure of X-ray luminosity density (or volume
emissivity) of the local Seyfert~2 population as a function of photon energy.
By integrating the luminosity density function over redshift (see Moran et
al.\ 2001), we obtain an estimate of the total X-ray sky brightness 
(i.e., XRB contribution) associated with Seyfert~2 galaxies.  
Combining our results with those of the {\em ROSAT\/}
Deep Survey (Schmidt et al.\ 1998), which established that $\sim 60$\% of the
XRB at 1~keV is produced by steep-spectrum type~1 AGN, we find that the
combined emission of Seyfert~1 and Seyfert~2 galaxies provides an 
excellent match to the spectrum and intensity of the XRB in the 
$2-10$~keV range (see Fig.~\ref{moranfig2}).  

Our approach is almost entirely empirical: the only model dependent
feature (and, thus, the main uncertainty) concerns the redshift evolution of
the AGN X-ray luminosity function (XLF).  In our study, we assume the evolution
of the Seyfert~2 XLF to be identical to that obtained for type~1 AGN in the
soft X-ray band by Miyaji et al.\ (2000). Updating our
calculations with an evolution term derived from {\em Chandra\/} and
{\em XMM-Newton\/} data (e.g., Cowie et al.\ 2003; Steffen et al.\ 2003;
Ueda et al.\ 2003; Hasinger et al.\ 2003) might change in detail the 
XRB estimates we 
obtain, but the qualitative agreement between our results and AGN models 
for the XRB would persist: (1) Seyfert~2 galaxies are the only known 
sources that, as a class, have the X-ray spectral properties needed 
to explain the shape of the XRB spectrum, and (2) their luminosity 
density is sufficient to account for the intensity of the hard XRB.  
In other words, there is very little room for a significant XRB 
contribution by some new population of X-ray--bright sources.

%
%

Locally, the optical spectra of Seyfert~2
galaxies are characterized by strong, narrow emission lines; spectroscopy of
the distant sources responsible for most of the hard X-ray flux in deep
{\it Chandra\/} images should, therefore, provide a straightforward means
of confirming the Seyfert~2 model for the XRB.  But a different picture has
emerged.  Follow-up observations of faint {\it Chandra\/} sources reveal
instead a significant population of apparently {\it normal\/} galaxies whose
starlight-dominated optical spectra have only weak emission lines, if any
(e.g., Mushotzky et al.\ 2000; Barger et al.\ 2001a, 2001b, 2002).  Many
such sources have the X-ray properties of Seyfert~2 galaxies, but they seem
to lack the associated optical emission-line signatures.  This could be taken
as evidence that the optical properties of hard X-ray galaxies have evolved
strongly with cosmic time, which would have important implications for the
nature of supermassive black holes and their environments at earlier epochs.
However, before settling in on this conclusion, it is imperative that we
first exhaust all possibility that distant hard X-ray galaxies are, in fact,
fundamentally similar to the familiar sources we find locally.  
Our attempts to do so are described in the next section.

\section{Comparing Local Seyfert~2 Galaxies and Distant X-ray Galaxies}

\subsection{Observational Challenges}

Any investigation of the properties of distant X-ray galaxies is 
naturally done within the context of our understanding of the local 
population. However, comparisons of nearby and distant sources can 
be complicated by two important factors.  First, samples of local 
and high-redshift X-ray galaxies are likely to be defined in very 
different ways.  Nearby X-ray galaxy samples can be assembled on the 
basis of a wide variety of criteria, including (but not limited to) 
X-ray brightness, strength of their emission in some other region
of the spectrum (e.g., the infrared), optical emission-line properties, 
or distance from the Milky Way.  Distant X-ray galaxies are usually 
identified on the basis of a sole property: detection as an X-ray 
source.  Thus, nearby and distant samples may contain inherently 
different types of X-ray galaxies and/or similar galaxies drawn from 
very different portions of the XLF.

As a second complicating factor, the quality of data typically 
available for nearby and distant X-ray galaxies differs vastly.  
Because of their proximity, the properties of nearby sources are 
often well characterized over a range of wavelengths.  For example, 
data for local active galaxies might include detailed broadband 
X-ray spectra, high signal-to-noise ratio (S/N) optical spectra 
of their nuclei, and robust detections in the radio, infrared, and
ultraviolet bands.  In contrast, faint X-ray galaxies at moderate 
redshifts might be characterized by the detection of a few tens 
of X-ray photons and low S/N or low resolution {\em integrated\/} 
optical spectra (or no spectra at all---optical photometry in a 
few bands may be all that is available); they are often unobserved 
or undetected at all other wavelengths.  Obviously, comparisons of 
sources with essentially incomparable data must be carried out with 
great care.

\subsection{X-ray--to--Optical Flux Ratios of the Nearby and Distant
Populations}

Distant, hard X-ray galaxies tend to be faint at all wavelengths, 
which limits the amount and quality of information we have about 
their properties.  For example, over half of the X-ray sources 
detected in the 2~Ms {\em Chandra\/} Deep Field-North (CDF-N; 
Alexander et al.\ 2003; Barger et al.\ 2003) have optical counterparts 
that are fainter than $R = 23$.  Clearly, high-quality optical 
spectra can only be obtained for the small fraction of relatively
bright sources included in that survey.  However, broadband 
magnitudes and colors have been measured for nearly all of the 
CDF-N sources.  One of the best observational handles we have on 
the nature of distant X-ray galaxies, therefore, is their 
X-ray--to--optical flux ratios, $F_X/F_{opt}$.
The $F_X/F_{opt}$ ratio broadly discriminates between 
various classes of celestial X-ray sources (e.g., Stocke et al.\ 1991);
in particular, between luminous AGN and ``normal'' (i.e., quiescent or 
non-active) galaxies.  Thus, one way to explore the nature of the 
optically-normal hard X-ray galaxies that are turning up in deep 
X-ray surveys would be to compare their $F_X/F_{opt}$ 
ratios to those of local X-ray galaxies with similar high-energy 
properties.  The results of such a comparison, carried out by
Moran \& Cardamone (2004), are summarized here.

To ensure that our flux ratio comparison is valid, we require 
an appropriate sample of distant X-ray galaxies from a 
well-characterized deep survey, and an unbiased sample of nearby 
sources with broadband X-ray and optical data. For the distant 
X-ray galaxy sample, the 2~Ms CDF-N is an ideal resource.  The
details of the {\em Chandra\/} observations and parameters of the 
detected sources are thoroughly documented (Alexander et al.\ 2003), 
and deep optical imaging and spectroscopy of the sources have 
been obtained with the Subaru 8~m and Keck 10~m telescopes 
(Barger et al.\ 2002, 2003).  From the CDF-N, we select only 
sources with total exposure times between 1.5~Ms and 2.0~Ms;
this range brackets the strong peak in the CDF-N source 
exposure time distribution centered at 1.7~Ms (Alexander et al.\ 2003), 
and because it is narrow, it allows us to establish an effective 
X-ray flux limit and solid angle for this portion of the deep survey.

Next, since we are chiefly concerned with the origin of the XRB, 
we select CDF-N sources with $2-8$~keV hard-band detections and 
flattened X-ray spectra with effective photon indices 
$\Gamma<1.5$ (as indicated by their ``hardness ratios''). These 
are the sources responsible for the hard XRB, and based on 
observations of nearby sources, they are expected to be Seyfert~2 
galaxies. Finally, we require that the included sources have a 
measured spectroscopic redshift.  About 60 CDF-N sources satisfy 
all of these criteria.  Using published $2-8$~keV fluxes and 
$I$-band magnitudes, we have computed the observed-frame 
$F_X/F_I$ flux ratios for these sources.

%
%
\begin{figure}[tbh]
\centerline{\psfig{figure=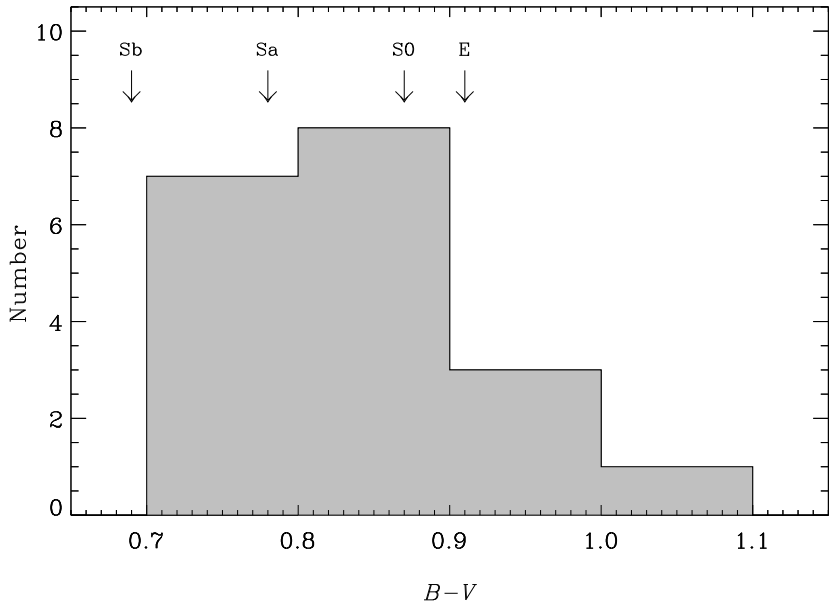,width=11truecm,angle=0}}
\caption{$B-V$ colors for the UW89 Seyfert~2 galaxies included in the 
RC3 (de Vaucouleurs et al.\ 1991).  Arrows indicate the typical colors 
of galaxies of various morphological types (Cox 2000).  The UW89 
galaxies tend to have red colors similar to the colors of 
early-type spirals. This is consistent with their morphological 
classifications, which are predominantly Sa, S0/a, or S0.
}
\label{moranfig3}
\end{figure}

The local Seyfert~2 galaxies for our study are again selected from 
the distance-limited sample defined by UW89.  As mentioned above, 
broadband X-ray observations of these sources are nearly complete, 
and because of their proximity, optical photometric data for the 
galaxies are available in the literature.  About two-thirds of the 
sources have integrated $UBV$ magnitudes from the 
Third Reference Catalog of Bright Galaxies (RC3;
de Vaucouleurs et al.\ 1991). Their $B-V$ colors are displayed in 
Figure~\ref{moranfig3}, along with an indication of the typical colors 
of various types of galaxies. As the figure illustrates, the host 
galaxies of nearby Seyfert~2 galaxies have the colors of early-type 
spirals---consistent with the morphological classifications
of the UW89 sources, which are predominantly Sa, S0/a, or S0.  
Interestingly, the distant, ``normal'' X-ray galaxies detected in 
deep {\em Chandra\/} images tend to have red colors and early-type 
optical spectra.

Since the $F_X/F_{opt}$ ratio is measured in the observed 
frame, its value depends to some degree on the redshift of an object. 
This redshift dependence, combined with differences in the ways 
samples of nearby and distant sources are assembled, makes a direct 
comparison of their flux ratio distributions impossible. We return 
to this point below.  The proper approach, therefore, involves a 
determination of the flux ratios that the nearby sources would have
if they were observed under the same conditions as the CDF-N sources,
and with the same redshift distribution as the CDF-N sources selected 
above.

%
%
\begin{figure}[tbh]
\centerline{\psfig{figure=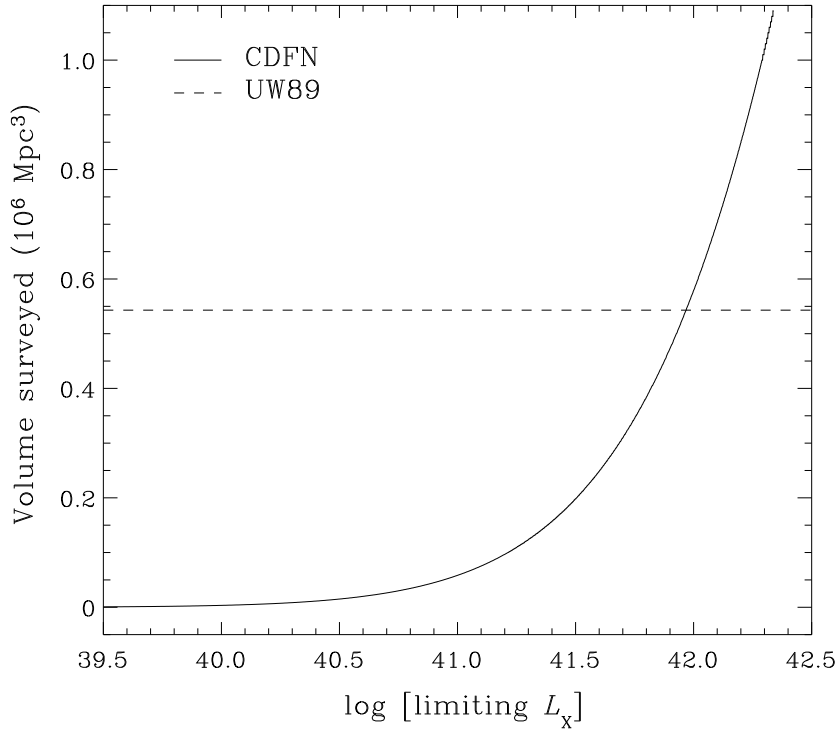,width=11truecm,angle=0}}
\caption{Volume searched in the 2~Ms {\em Chandra} Deep Field-North,
as a function of minimum detectable $2-8$~keV X-ray luminosity
$L_X$ {\em (solid curve)}. 
Also plotted is the volume covered by the UW89 sample
of nearby Seyfert~2 galaxies {\em (dashed line)}.  
An $\Omega_{\rm tot} = 1$,
$\Omega_{\rm M} = 1/3$, $\Omega_{\Lambda} = 2/3$ cosmology with
a Hubble constant of $H_0 = 70$~km~s$^{-1}$~Mpc$^{-1}$ has been
assumed.
}
\label{moranfig4}
\end{figure}

We use Monte Carlo methods to simulate the $F_X/F_I$ distribution of 
the local Seyfert~2 galaxies, randomly selecting a redshift 
(weighted by the CDF-N redshift distribution) and a UW89 galaxy 
(unweighted, since to first order the UW89 sample {\em is\/} the 
local Seyfert~2 luminosity function).  We begin by using the 
{\em ASCA\/} data to calculate the observed-frame $2-8$~keV flux 
of the selected galaxy, verifying that it would exceed the CDF-N 
flux limit.  Next, we determine the likelihood that a UW89 galaxy 
of a particular luminosity would be included in the CDF-N.  For 
this test, we have combined the CDF-N flux limit and survey
solid angle to estimate the volume searched in the CDF-N as a 
function of minimum detectable hard X-ray luminosity $L_X$.  
The results are plotted in Figure~\ref{moranfig4}, along with 
the (constant) volume covered by the UW89 sample.  Below
a $2-8$~keV luminosity of $\sim 10^{42}$ ergs~s$^{-1}$, the 
volume searched in the CDF-N is less than that of the UW89 sample.  
Therefore, in this $L_X$ range, the ratio of the CDF-N volume 
to the UW89 volume defines the probability that a local object 
of a given luminosity would be detected in the CDF-N.  If
a galaxy passes all tests, its $UBVRI$ magnitudes are used to 
obtain its observed-frame $I$-band flux, from which its 
$F_X/F_I$ ratio is calculated.

%
%
\begin{figure}[tbh]
\centerline{\psfig{figure=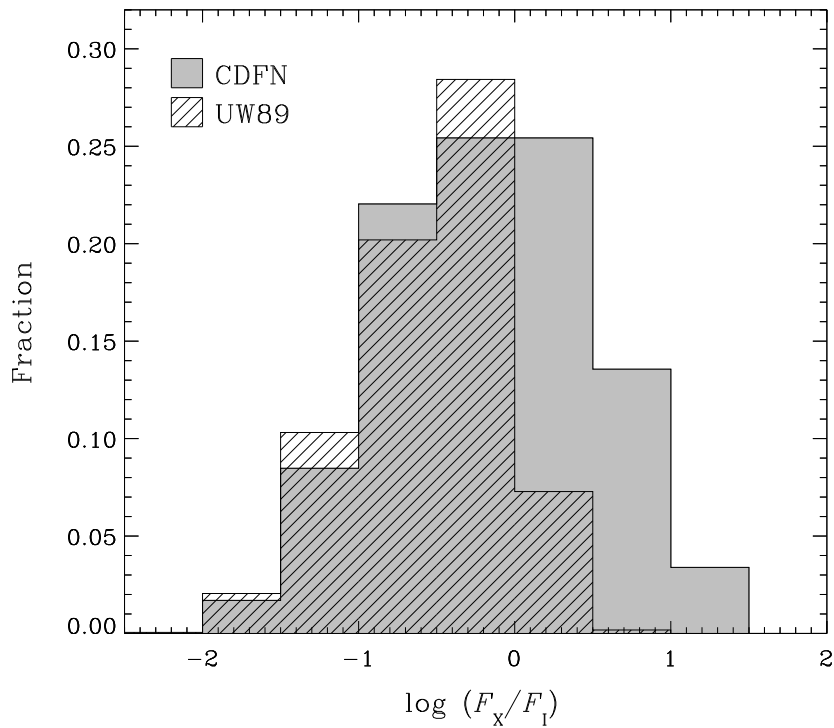,width=11truecm,angle=0}}
\caption{Distribution of the observed-frame $2-8$~keV/$I$-band flux
ratio for CDF-N sources with Seyfert~2-like X-ray properties,
compared to the $F_X/F_I$ ratios that UW89 Seyfert~2 galaxies
would have if they were observed in the CDF-N. The UW89 distribution
has been normalized to have the same area as the the CDF-N distribution
in the $\log F_X/F_I = -1.5$ to $-0.5$ range, to emphasize
the similarity between the two distributions at low values of the
flux ratio.
}
\label{moranfig5}
\end{figure}

In Figure~\ref{moranfig5}, we compare the $F_X/F_I$ 
distribution of the hard XRB producing CDF-N sources to the 
simulated flux ratio distribution that the UW89 Seyfert~2 galaxies would 
have if they were observed in the CDF-N.  Below 
$F_X/F_I \approx 0$, the two distributions match 
closely---I would argue that the minor differences present are 
mainly due to the discreteness of the UW89 sample.  However, at 
the higher values of $F_X/F_I$, only CDF-N sources 
are present.  Is this an indication of some fundamental difference 
between the nearby and distant populations?

%
%
\begin{figure}[tbh]
\centerline{\psfig{figure=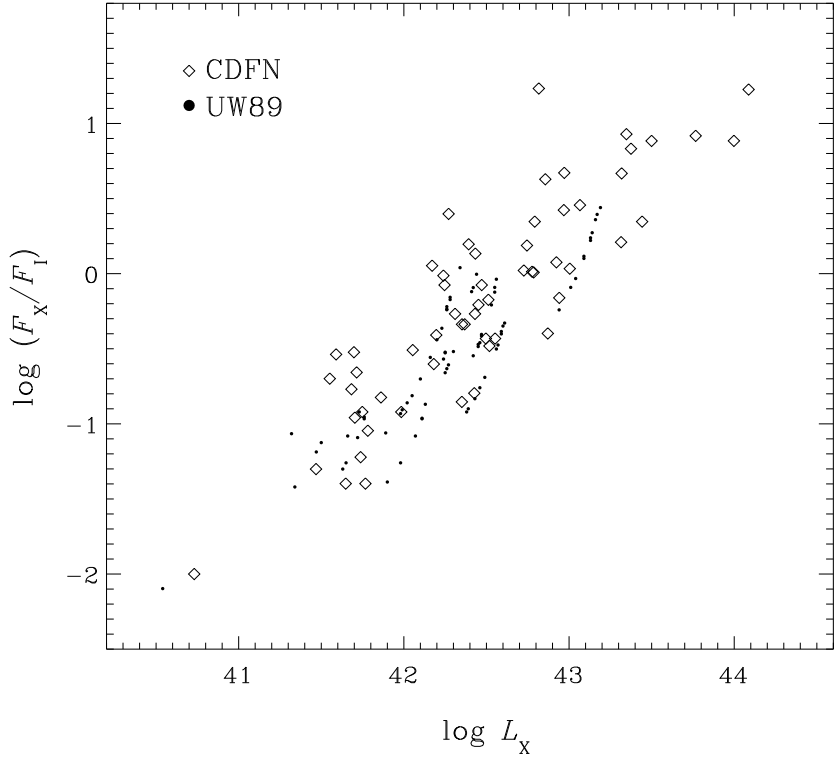,width=11truecm,angle=0}}
\caption{$F_X/F_I$ ratio for the CDF-N and UW89 samples,
as a function of observed hard X-ray luminosity. There is excellent
overlap between the two sets of points below
$L_X\approx 10^{43}$~ergs~s$^{-1}$.
}
\label{moranfig6}
\end{figure}

Some clarification is provided by Figure~\ref{moranfig6}, where 
$F_X/F_I$ is plotted as a function of hard X-ray 
luminosity $L_X$.  The observed correlation between these 
two quantities suggests that the dispersion in the optical fluxes 
of the galaxies is not too great.  Below $L_X \approx
10^{43}$ ergs~s$^{-1}$, the CDF-N points and the UW89 points from 
the simulation overlap quite nicely.  Note that high values of 
$F_X/F_I$, where UW89 points are absent, correspond 
to high values of $L_X$. Referring once more to 
Figure~\ref{moranfig4}, the volume searched in the CDF-N for
high $L_X$ sources vastly exceeds that covered by the UW89 
sample. Thus, the apparent differences in the $F_X/F_I$ 
distributions are likely to be 
the result of Malmquist effects: the volume associated
with the UW89 sample is too small to include rare, high-luminosity 
sources, which are over-represented in the CDF-N, because of the 
large volume it surveys for such sources.  We conclude, therefore, 
that the $F_X/F_I$ distributions of distant, 
flat spectrum X-ray galaxies and nearby Seyfert~2 galaxies
do {\em not\/} differ appreciably, at least for sources with 
observed hard X-ray luminosities below a few times 
$10^{43}$ ergs~s$^{-1}$.  If the Malmquist effects described 
above are present---and to some extent they must be---this 
conclusion may be valid over all values of $L_X$.

%
%
\begin{figure}[tbh]
\centerline{\psfig{figure=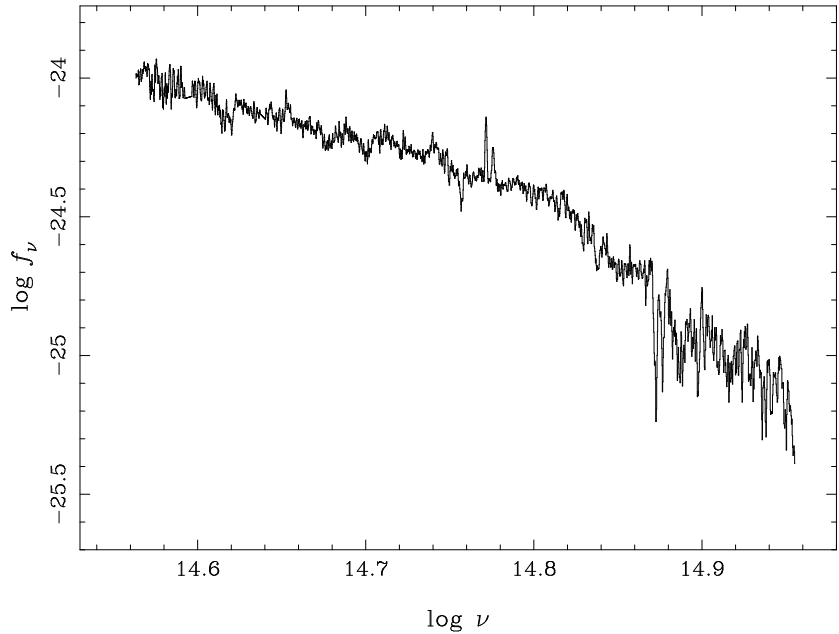,width=11truecm,angle=270}}
\caption{Optical spectrum of the UW89 Seyfert~2 galaxy NGC~788.
The part of the spectrum that falls within the rest-frame $I$-band
is that below $\log \nu = 14.63$.
}
\label{moranfig7}
\end{figure}

This exercise has lead to some interesting insights into the way 
source redshifts and the flux-limited nature of deep surveys combine 
to influence the Seyfert~2 $F_X/F_I$ distribution.  
For Seyfert~2 galaxies, the redshift dependence of the X-ray--to--optical 
flux ratio is quite strong. Plotted in Figure~\ref{moranfig7} is 
the integrated optical spectrum of NGC~788, one of the more luminous 
X-ray sources in the UW89 sample.  The broadband X-ray spectrum of 
NGC~788 is very similar to the composite Seyfert~2 spectrum
displayed in Figure~\ref{moranfig1}.  As photon energy (or frequency) 
increases, the X-ray spectrum of NGC~788 rises, and its optical 
spectrum declines.  Thus, the observed-frame $F_X/F_I$ 
ratio of this source must increase with redshift.

%
%
\begin{figure}[tbh]
\centerline{\psfig{figure=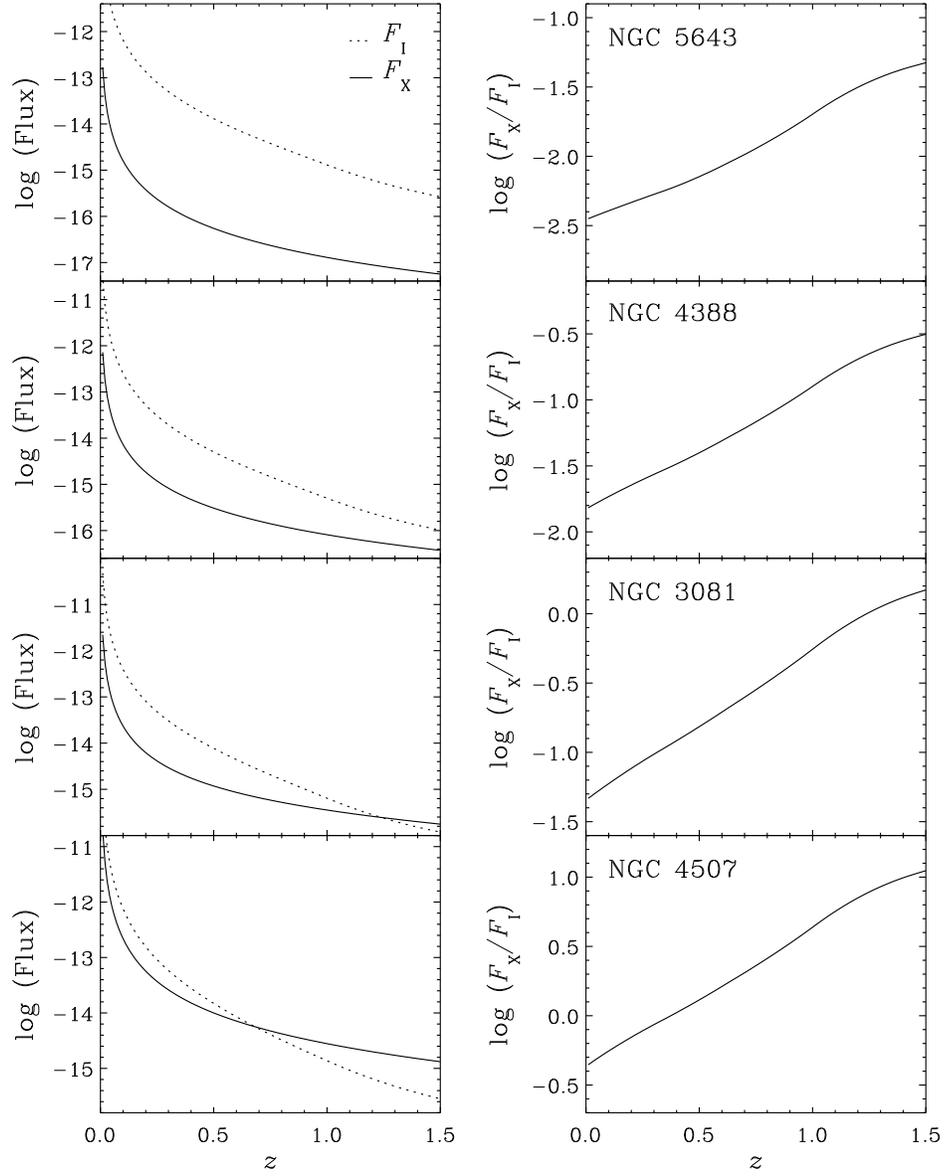,width=12.5truecm,angle=0}}
\caption{{\em (Left)} Observed-frame $2-8$~keV {\em (solid curve)} and 
$I$-band flux {\em (dashed curve)} vs. redshift, and {\em (right)} 
their ratio vs. redshift, for four UW89 Seyfert~2
galaxies that span a wide range of $F_X/F_I$ at $z = 0$.  The
$F_X/F_I$ ratio increases dramatically for all four sources
as $z$ increases.
}
\vskip 0.6cm
\label{moranfig8}
\end{figure}

\begin{figure}[tbh]
\centerline{\psfig{figure=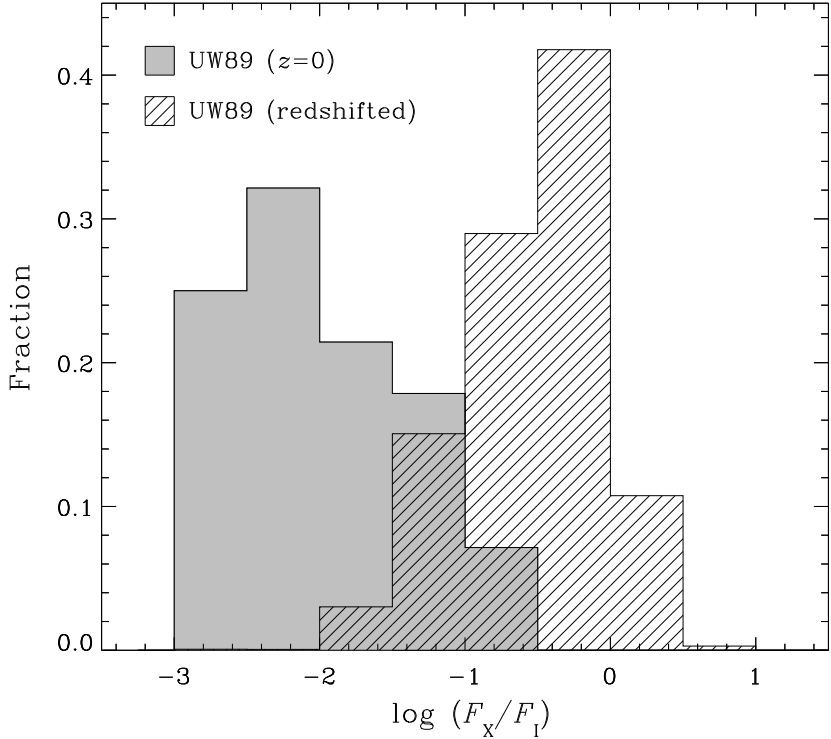,width=11truecm,angle=0}}
\caption{Intrinsic ($z = 0$) $F_X/F_I$ distribution for
the UW89 sample {\em (shaded histogram)}, 
and the distribution that would be obtained if the same
sources were observed in the CDF-N {\em (hatched histogram)}.  
The dramatic transformation of 
the flux ratio distribution is a combination of redshift effects and 
the bias towards high-luminosity (thus high $F_X/F_I$) 
sources that results from the flux-limited nature of the CDF-N.
}
\label{moranfig9}
\end{figure}

In Figure~\ref{moranfig8}, we plot X-ray and optical fluxes and 
their ratio versus redshift for four UW89 Seyfert~2 galaxies that span a 
wide range of intrinsic (i.e., $z = 0$) values of $F_X/F_I$.  
At $z = 1.5$, their observed-frame X-ray--to--optical flux ratios are 
increased by factors of 15 to 30.  Combined with the effects 
associated with the flux-limited nature of the CDF-N (i.e., the 
suppression of the faint end of the Seyfert~2 XLF, and the 
over-representation of sources at the bright end), redshift 
effects cause a dramatic transformation of the distribution of 
$F_X/F_I$ for Seyfert~2 galaxies, as illustrated in 
Figure~\ref{moranfig9}.

%
%

\subsection{Integrated Spectra of Seyfert~2 Galaxies}

If a careful comparison of the $F_X/F_{opt}$ ratios of 
nearby Seyfert~2 galaxies and distant hard X-ray galaxies fails to reveal 
significant differences between the two populations, why do their 
optical spectra differ? One possibility is that AGN in the past 
tend to be even more obscured than Seyfert~2 galaxies.  If so, they might 
exhibit none of the optical properties displayed by nearly all 
local X-ray luminous AGN.  As an alternative to the strong AGN 
evolution implied in this scenario, we propose that
the optical results of the deep {\em Chandra\/} surveys may instead 
be heavily influenced by the limitations of ground-based observing.  

In spectroscopic observations of nearby AGN, light is collected 
through a small aperture centered on the nucleus, which excludes 
most of the starlight from the host galaxy.  The faint sources 
that produce the hard XRB, in contrast, are typically so distant 
that when they are observed, the entire galaxy (or a large fraction 
of it) falls within the spectrograph slit.  (At $z \approx 1$,
10 kpc corresponds to $\sim$~1\farcs5.)  Combined with the low 
spectral resolution that has been employed to date ($\sim 12-20$~\AA) 
and the modest signal-to-noise ratios frequently obtained (because 
the optical counterparts are so faint), the additional galaxy 
light from stars and \HII\ regions could lead to the appearance 
that some distant {\em Chandra\/} sources are associated with normal 
galaxies rather than with Seyfert~2 galaxies.

\begin{figure}[tbh]
\centerline{\psfig{figure=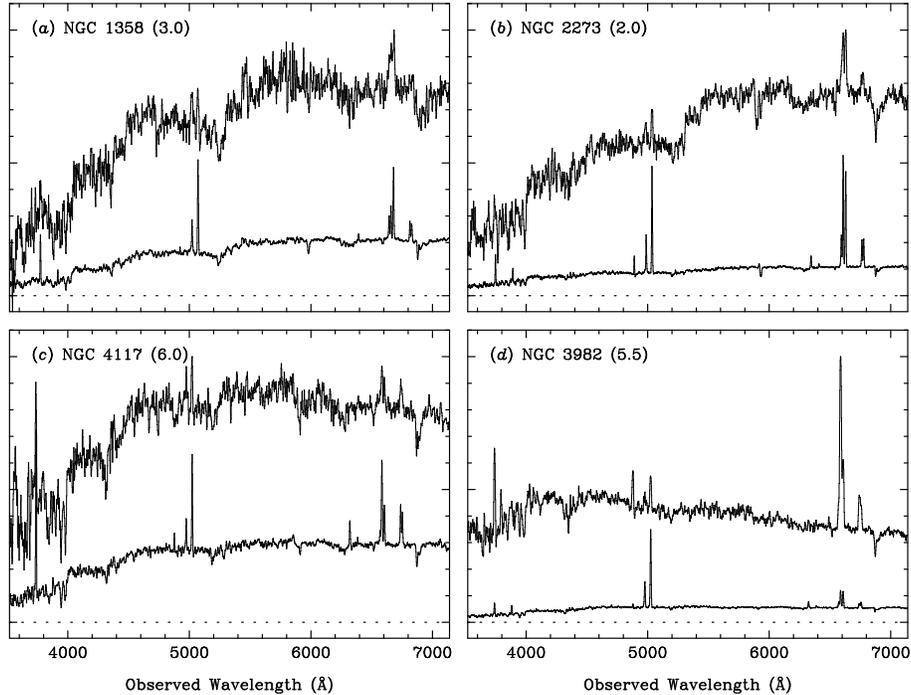,width=12truecm,angle=0}}
\caption{Data for four UW89 Seyfert~2 galaxies observed by 
Moran et al.\ (2002).  Both nuclear {\em (lower trace)} and 
integrated {\em (upper trace)} spectra are shown in each panel.  
Relative flux densities (in $f_{\lambda}$ units) are plotted on 
the ordinate.  For clarity, the nuclear spectrum of each object
has been multiplied by a constant, which appears following the 
galaxy name. In all four cases, the nuclear emission lines are 
far less prominent in the integrated spectrum, which provides 
strong support for the hypothesis that host galaxy dilution is 
responsible for the normal appearance of many distant galaxies 
discovered in hard X-ray surveys.
}
\label{moranfig10}
\end{figure}

To investigate this possibility, we have obtained {\em integrated\/} 
spectra of 18 nearby Seyfert~2 galaxies that are known to be 
absorbed X-ray sources (Moran, Filippenko, \& Chornock 2002).  
The sources for this work were again selected from the UW89 sample.  
Our techniques simulate spectroscopic observations of distant X-ray 
galaxies with Keck, which allow us to evaluate whether the 
emission-line signatures of their activity can be overwhelmed
(or ``hidden'') in the spectra of their integrated light.  The 
results are striking: as Figure~\ref{moranfig10} shows, the nuclear 
emission lines of many sources are almost completely washed out in 
the integrated spectra.  As with most of the normal looking 
{\em Chandra\/} sources, weak emission lines are present in
three of the integrated spectra in Figure~\ref{moranfig10}, 
but the strongest features are the \CaII , G band, \MgIb, and 
\NaID\ stellar absorption lines typically observed in the spectra 
of inactive early-type galaxies.  Emission lines are quite strong 
in the integrated spectrum of the fourth object
(Fig.~\ref{moranfig10}$d$), but the line-intensity ratios---similar 
to the values observed in \HII\ regions---are drastically different 
than those observed in the nucleus.  At a modest redshift, this 
object would be classified as a starburst galaxy!

%
%

About 60\% of the 18 nearby sources we observed would not be 
classified as AGN on the basis of their integrated spectra. 
We conclude that it {\em is\/} possible to hide the true nature 
of a significant fraction of distant type~2 AGN in their 
integrated optical spectra.  Spectral coverage issues may affect
the deep survey results as well.  The classifications of some of 
the persistent Seyfert~2 galaxies in our sample would be ambiguous 
without information about the emission lines near \Ha .  Because of 
their high redshifts, many CDF-N galaxies have spectra that do not 
cover the \Ha\ region.  Thus, the combined limitations of ground-based 
observations---not some sort of evolutionary effects---may be 
largely responsible for the current demographics of the distant 
hard X-ray galaxy population.

\section{Summary and Future Work}

Spectroscopic studies of faint X-ray sources in deep {\em Chandra\/} 
images have apparently uncovered a significant population of distant, 
optically-normal galaxies.  These sources have X-ray properties 
similar to those of nearby Seyfert~2 galaxies, but they seem to 
lack the strong optical emission lines that characterize the latter.  
This suggests that the X-ray galaxy population has perhaps undergone 
significant optical evolution, which has stimulated efforts to improve 
our understanding of X-ray--bright, optically-normal galaxies 
(e.g., Comastri et al.\ 2002) and searches for examples of
them in the local universe (Maiolino et al.\ 2003).

While the prospect of strong evolution of the X-ray galaxy population 
is tantalizing, it is important to explore the possibility that the 
faint, distant sources responsible for most of the hard XRB are, 
in fact, optically similar to Seyfert~2 galaxies.  Ground-based 
optical spectra of deep survey sources include not only the nuclear 
emission we wish to relate to their X-ray properties, but the 
majority of the light from stars and \HII\ regions in the host 
galaxies, as well.  In some cases, the extranuclear host galaxy
light might overwhelm the nuclear emission, giving the appearance that 
an absorbed AGN is an inactive galaxy.  

To test this hypothesis, we have obtained integrated optical 
spectra of nearby Seyfert~2 galaxies. We find that
about 60\% of such sources lack clear signs of nuclear activity in 
the spectra of their integrated light.  Moreover, the observed-frame 
X-ray--to--optical flux ratios of distant hard X-ray galaxies, and 
those that nearby Seyfert~2 galaxies would have if they were observed 
in an identical manner, are very similar.  Thus, there is insufficient 
evidence at this time to support a conclusion that distant hard X-ray 
galaxies differ significantly from nearby Seyfert~2 galaxies.

Of course, the tests presented here are indirect, and they do not 
rule out all possibility that distant hard X-ray galaxies have evolved.  
High angular resolution optical spectroscopy, isolating the nuclei 
of such sources, would help to settle the evolution/starlight dilution 
debate directly.  My colleagues and I are preparing to do just that---we 
have an approved {\em Hubble Space Telescope\/} program to obtain 
spectra of four sources with the Space Telescope Imaging Spectrograph 
(STIS).  The targets to be observed have normal looking ground-based 
optical spectra, but hard X-ray luminosities in excess of 
$10^{42}$ ergs~s$^{-1}$, too high to be associated with 
truly normal galaxies.  The redshifts of the sources are
in the $z\approx 0.2$ range, so spectra of their nuclei with the 
0\farcs1 or 0\farcs2 STIS slits will exclude the vast majority of 
the emission from the host galaxy.  The results of the observations
will provide much needed clarification of the properties of distant 
hard X-ray galaxies and, thus, valuable insight into the nature 
of the supermassive black holes they contain.

\begin{acknowledgments}
I am grateful to John Salzer for insightful discussions about Malmquist
effects in flux-limited surveys, and to Carolin Cardamone for assistance
with many of the calculations and figures presented here.
\end{acknowledgments}

\end{document}